\documentclass[prc,aps,twocolumn,floatfix,unsortedaddress,reprint]{revtex4-1}

\usepackage{graphics}
\usepackage{epsfig}
\usepackage{amsmath}
\usepackage{amssymb}
\usepackage{lineno}

\begin{document}

\title{Gamow-Teller transitions and the spin EMC effect: \\
the Bjorken sum-rule in medium}

\author{Steven D. Bass}
\email{Steven.Bass@cern.ch}
\affiliation{\mbox{Kitzb\"uhel Centre for Physics,
Kitzb\"uhel, Austria}}
\affiliation{\mbox{Marian Smoluchowski Institute of Physics, Jagiellonian University, 
PL 30-348 Krakow, Poland}}

\begin{abstract}
Gamow-Teller transitions in nuclei tell us that the nucleon's
axial charge $g_A^{(3)}$ is quenched in large nuclei by about 20\%.
This result tells us that the spin structure of the nucleon
is modified in nuclei
and disfavours models of the medium dependence of 
parton structure 
based only on nucleon short range correlations in nuclei.
For polarized photoproduction 
the Gerasimov-Drell-Hearn integral is expected to be strongly
enhanced in medium.
\end{abstract}

\maketitle

\section{Introduction}

Just about 30\% of the proton's spin is carried by the spin 
of its quarks. 
This surprising discovery 
from polarized deep inelastic scattering
has inspired a 30+ years global programme of theory and experiments
to understand the internal spin structure of the proton
\cite{Aidala:2012mv,Bass:2004xa}.
In parallel,
unpolarized deep inelastic scattering from nuclear targets 
has taught us that the 
quark structure of the proton 
is modified when the proton is inside an atomic nucleus.
Detailed explanation of this EMC nuclear effect
is still a matter of theoretical debate, 
for recent discussion see \cite{Cloet:2019mql}.
New experiments are planned at Jefferson Laboratory with a 
polarized $^7$Li target to look for a possible spin version
of the EMC nuclear effect 
in the range $0.06 < x < 0.8$ \cite{JLabexpt}.
How is the internal spin structure of the proton 
modified when the proton is in a nuclear medium?

Here we explain how Gamow-Teller transitions 
($\beta$-decays of large nuclei) 
constrain our understanding of nucleon spin structure in medium and models of the EMC nuclear effect. 
The effective 
isovector axial charge $g_A^{(3)}$ 
extracted from Gamow-Teller transition experiments 
is quenched in large nuclei by about 20\% \cite{ericson}.
Through the Bjorken sum-rule \cite{Bjorken:1966jh,Bjorken:1969mm}, 
this means a corresponding reduction in the difference between 
up and down quark spin contributions to the proton's spin in
the nuclear medium.
The dynamics of this quenching of $g_A^{(3)}$ is
driven by combination of 
the Ericson-Ericson-Lorentz-Lorenz effect and
pion cloud effects in nuclei
with short range nucleon nucleon correlations tending
to reduce the axial-charge suppression
 \cite{Ericson:1998hq,Ericson:1973vj}.
This result has
important consequences for models of the EMC nuclear effect.
Specific models of partonic spin structure in medium 
have been discussed
in Refs.~\cite{deBarbaro:1984gh,Szwed:1985vxp,Guzey:1999rq,Sobczyk:2000rf,Cloet:2005rt,Smith:2005ra,Cloet:2006bq,Tronchin:2018mvu,Thomas:2018kcx}.

Popular models of the EMC nuclear effect
involve either modification of the properties of each nucleon
in the nucleus through coupling of the valence quarks
to the scalar and vector mean fields in the nucleus
or
where most nucleons are unmodified 
but a small number exist in short range correlations
where the struck nucleon 
is far off mass shell \cite{Cloet:2019mql}.
Models of the EMC nuclear effect where the effect is driven
only by nucleon short range correlations in nuclei
predict a negligible spin effect in medium \cite{Thomas:2018kcx}, 
in contrast to the phenomenological constraint 
from the quenching found in Gamow-Teller transitions.

In Section II we give a brief overview of present
understanding of the proton's spin structure.
Section III discusses the constraints from medium modifications
of $g_A^{(3)}$.
In Section IV we discuss the consequences for models of the
EMC nuclear effect and outlook for future experiments. 
Section V addresses the extension to polarized photoproduction
where 
the value of the
Gerasimov-Drell-Hearn, GDH, sum-rule 
is expected to be strongly enhanced in medium.
Conclusions are given in Section VI.

\section{The spin structure of the proton in free space}

Information about the proton's spin structure comes 
from the $g_1$ deep inelastic spin structure function. 
In QCD the first moment of $g_1$ is given by a linear
combination of the nucleon's isovector, octet and 
flavour-singlet axial charges, each times perturbative 
QCD coefficients which are calculated to $O(\alpha_s^3)$ precision. 
For quark flavour $q$, the axial-charges
\begin{equation}
2M S_{\mu} \Delta q =
\langle p,S |
{\overline q} \gamma_{\mu} \gamma_5 q
| p,S \rangle
\label{eqc55}
\end{equation}
measure the fraction of the proton's spin that 
is carried by quarks and antiquarks of flavour $q$.
Here $M$ is the proton's mass and $S$ its spin vector.
The isovector, octet and singlet axial charges are
\begin{eqnarray}
g_A^{(3)} &=& \Delta u - \Delta d
\nonumber \\
g_A^{(8)} &=& \Delta u + \Delta d - 2 \Delta s
\nonumber \\
g_A^{(0)}
&=& \Delta u + \Delta d + \Delta s
.
\label{eqc56}
\end{eqnarray}
Each spin term $\Delta q$ 
($q= u,d,s$)
is understood to contain a contribution from polarized gluons,
$- \frac{\alpha_s}{2 \pi} \Delta g$,
where $\alpha_s$ is the QCD coupling 
and 
$\Delta g$ is the polarized gluon contribution
to the proton's spin.
This polarized gluon term contributes in 
$g_A^{(0)}$ but cancels in $g_A^{(3)}$ and $g_A^{(8)}$.
The value of the singlet 
$g_A^{(0)}$ is also sensitive to a 
possible topological contribution, ${\cal C}_{\infty}$
which, if finite, 
is associated with Bjorken $x=0$ and a 
subtraction constant from the 
``circle at infinity'' in the dispersion relation for $g_1$
\cite{Bass:2004xa}.

For free protons,
in QCD the isovector part of $g_1$ satisfies the fundamental
Bjorken sum-rule
\begin{equation}
\int_0^1 dx g_1^{(p-n)} (x,Q^2) 
= \frac{g_A^{(3)}}{6} C_{\rm NS} (Q^2) 
\end{equation}
where 
$x$ is the Bjorken variable,
$g_A^{(3)} = 1.270 \pm 0.003$  
from neutron $\beta$-decays and
$C_{NS}(Q^2)$
is the perturbative QCD Wilson coefficient, $\simeq 0.85$ 
with QCD coupling $\alpha_s = 0.3$~\cite{Aidala:2012mv}.
This sum-rule has been confirmed in polarized deep inelastic scattering experiments at the level of 5\% 
\cite{Alekseev:2010hc}.
About 50\% of the sum-rule comes from Bjorken $x$ 
values less than about 0.15.
The $g_1^{(p-n)}$ data is consistent with quark model and 
perturbative QCD predictions in the valence region $x > 0.2$ \cite{Bass:1999uj}.
The size of $g_A^{(3)}$ forces us 
to accept a large contribution from small $x$
with the observed rise 
\begin{equation}
g_1^{(p-n)} \sim x^{-0.22 \pm 0.07}
\end{equation}
found in COMPASS data from CERN at $Q^2 = 3$ GeV$^2$
for small $x$ data down to $x_{\rm min} \sim 0.004$
\cite{Alekseev:2010hc}.
Surprisingly, recent analysis \cite{Bass:2018uon}
of high statistics data
from the CLAS experiment at Jefferson Laboratory 
and COMPASS reveals that the 
rising behaviour in Eq.~(4)
persists to low $Q^2 < 0.5$ GeV$^2$
in contrast to the simplest 
Regge predictions based on a straight line $a_1$ trajectory.
This finding remains to be fully understood in terms of the
underlying QCD dynamics.
The effective Regge intercept $\alpha_{a_1} = 0.31 \pm 0.04$
\cite{Bass:2018uon}
gives the high energy part (about 10\%) of 
the Gerasimov-Drell-Hearn sum-rule for polarized
photoproduction which is needed to match on 
to low energy contributions measured at Bonn and Mainz
\cite{Helbing:2006zp}.

The isoscalar spin structure function
$g_1^{(p+n)} \sim 0$ for $x < 0.03$ at deep inelastic $Q^2$
\cite{Aidala:2012mv},
in sharp contrast to the unpolarized structure function $F_2$
where the isosinglet part dominates through gluonic exchanges.
The proton spin puzzle, why the quark spin content of 
the proton is so small $\sim 0.3$, concerns the 
collapse of the isoscalar spin sum structure function 
to near zero at this small $x$.
The spin puzzle involves contributions from the virtual
pion cloud of the proton
with transfer of quark spin to orbital angular momentum 
in 
the pion cloud 
\cite{Bass:2009ed}, 
the colour hyperfine interaction
or one-gluon-exchange current (OGE) \cite{Myhrer:2007cf},
a modest polarized gluon correction 
$- 3 \frac{\alpha_s}{2 \pi} \Delta g$ 
with $\Delta g$ 
non-zero \cite{deFlorian:2014yva}
and less than about 0.5 
at the scale of 
the experiments \cite{Aidala:2012mv}, 
and a possible topological effect at $x=0$ \cite{Bass:2004xa}.

\section{$g_A^{(3)}$ in medium}

Static properties of hadrons 
(masses, axial charges, magnetic moments...) 
are modified in a nuclear medium 
\cite{ericson,Bass:2018xmz,Metag:2017yuh,Saito:2005rv}.
For axial structure,
Gamow-Teller transitions ($\beta$ decays of large nuclei)
tell us that
the effective axial charge in medium $g_A^{* (3)}$ 
is suppressed in large nuclei by about 20\% \cite{ericson};
for recent reviews of experimental data see \cite{Suhonen:2018ykq,Suhonen:2017krv}.
This quenching is measured in the space component of 
the axial current with matrix element 
proportional to the nucleon spin vector $\vec{S}$.
Quenching of $g_A^{*(3)}$ in nuclei tells us that the 
spin structure of the nucleon is modified in nuclei
with
\begin{equation}
g_A^{*(3)} = \Delta u^* - \Delta d^* \simeq 1 
\end{equation}
close to nuclear matter density $\rho_0 = 0.15$ fm$^{-3}$
and with the Bjorken $x$ dependence of the effect waiting 
to be discovered.

Quenching of $g_A^{*(3)}$ can be understood in terms of
nucleon, $\Delta$ and pion degrees of freedom 
(without explicit quark and gluon degrees of freedom)
and
through coupling the valence quarks in the nucleon 
to the scalar and vector mean fields in the medium.
In the first approach,
important contributions 
come from the Ericson-Ericson-Lorentz-Lorenz effect
\cite{ericson,Ericson:1973vj}
and from interaction with the pion cloud in the nucleus
\cite{Ericson:1998hq}.
These terms each give about 50\% of the quenching effect.
Any contribution 
from short range nucleon correlations 
tends to reduce the quenching, 
see \cite{Ericson:1998hq} and Section IV below.
In a nuclear medium or nucleus relativistic invariance 
is lost and the space and time components of the axial 
vector current become disconnected.
Meson exchange currents provide extra renormalization of
the time component
of the axial current 
with enhancement seen in the time component in
$0^+ \leftrightarrow 0^-$ transitions,
in contrast to the quenching seen in the space component.
Chiral symmetry quenching effects are universal to the space and time components.

In a QCD motivated approach the quark meson coupling model, QMC,
predicts about 10\% reduction in $g_A^{*(3)}$ 
at $\rho_0$ \cite{Saito:1994kg}.
Here medium modifications of hadron properties
are calculated by treating the hadron as an MIT Bag and
coupling 
the valence quarks
to the scalar $\sigma$ (correlated two pion)
and
vector $\omega$ and $\rho$ mean fields in the nucleus.
Since one works in mean field there is no explicit 
Ericson-Ericson-Lorentz-Lorenz term in this model.
About 14\% reduction is found when OGE and pion cloud
effects are included in the model \cite{Nagai:2008ai}.
In recent QCD lattice calculations 
modest suppression of $g_A^{*(3)}$, a few percent,
is found for light nuclei \cite{Chang:2017eiq}.

What does the quenching of $g_A^{*(3)}$ mean for models of 
the EMC nuclear effect?

\section{Consequences for the EMC nuclear effect}

The EMC nuclear effect \cite{Cloet:2019mql}
involves suppression of
the unpolarized $F_2$ structure function in medium 
relative to the free nucleon structure function
in the valence region with 
Bjorken $x$ between about 0.3 and 0.85.
There is
enhancement around $x=0.15$,
the ratio comes with
constant negative slope between 0.15 and 0.7,
plus shadowing suppression at smaller $x$
which is expected to saturate at some small value of $x$
corresponding to $A$ independent effective Regge intercepts,
with $A$ the mass number.

What do we expect for spin?
The polarized EMC effect is defined through
\begin{equation}
\Delta R_A^H (x) =
\frac{g_1^{AH} (x)}{P_{AH}^p g_1^p(x) + P_{AH}^n g_1^n(x)}
\end{equation}
where $g_1^{AH}$ 
is the spin dependent structure function
for a nucleus with helicity $H$ and mass number $A$,
$g_1^p$ and $g_1^n$ are free nucleon structure functions and
$P_{AH}^p$ and $P_{AH}^n$ are the effective polarization of the protons and neutrons in the nucleus 
\cite{Sobczyk:2000rf,Cloet:2006bq}.

Today there are two leading approaches for describing 
the unpolarized EMC effect in the valence region.
Mean-field models have all of the nucleons slightly
modified through coupling their valence quarks to the scalar
and vector mean fields in the nucleus~\cite{Saito:2005rv}.
In a different view,
nucleons are unmodified most of the time but are modified substantially when they fluctuate into short range correlated pairs, SRCs \cite{Hen:2016kwk}.
Experimentally, a correlation is observed between SRCs
and the magnitude of the unpolarized EMC effect in nuclei
\cite{Weinstein:2010rt} raising the question whether
SRCs cause the EMC effect or whether both might have
a common origin so that one might have a spin EMC effect 
without SRCs having to induce it.

Early calculations in a model with explicit pion and 
$\Delta$ resonance degrees of freedom 
\cite{deBarbaro:1984gh,Szwed:1985vxp,Sobczyk:2000rf}
plus more recent 
calculations of the nucleon's $g_1$ spin structure
function in medium based on mean field approaches
\cite{Cloet:2005rt,Cloet:2006bq, Tronchin:2018mvu,Smith:2005ra}
suggest a large spin EMC effect in the valence region 
at medium $x$.
Calculations in the QMC model give a ratio of in-medium 
to free nucleon spin structure functions similar 
in size to the unpolarized EMC nuclear effect with 
$g_A^{*(3)}$ reduced by about 10\% at $\rho_0$
\cite{Tronchin:2018mvu}.
NJL model calculations give double the unpolarized effect
in the valence region with larger suppression of the 
ratio of spin structure functions for large nuclei
and $g_A^{*(3)}$ reduced by about 20\% at $\rho_0$, 
and with a constant EMC ratio $\Delta R_A^H(x) \sim 0.93$
for $x < 0.7$ with $g_A^{*(3)}$ reduced by about 6\% in 
$^7$Li~\cite{Cloet:2006bq}.
Shadowing at small $x$ is considered in
\cite{Guzey:1999rq}.

Models of the EMC nuclear effect where the effect 
is induced only by the contribution of short range 
nucleon correlations give only negligible spin dependence
\cite{Thomas:2018kcx}.
In SRCs 
two nucleons meet with low relative momentum and
relative angular momentum in $S-$wave.
Through the SRC
the nucleons will be scattered into a high
relative momentum $D-$wave state by the tensor force. 
Evaluating the relevant Clebsch-Gordon coefficients,
one finds that this process 
significantly depolarizes the correlated struck proton 
which is far off mass shell because of 
the high momentum carried away by its partner nucleon. 
The polarization of the struck nucleon participating 
in the SRC will be of order -10 to -15\% instead of +100\%
\cite{Thomas:2018kcx}.
That is, any medium modification induced by the SRC in 
the unpolarized structure function is washed out in the 
spin structure function $g_1^{AH}(x)$ and in 
$\Delta R_A^H(x)$. 
This contrasts with the 20\% quenching of
$g_A^{*(3)}$ expected from Gamow-Teller transitions.

In future experiments 
if no suppression is found in the valence region of the 
isovector part of $g_1$ in medium,
then 
$g_1^{(p-n)}$ in medium
should be strongly suppressed at smaller $x < 0.15$, 
where 50\% of
the Bjorken sum-rule for free protons comes from,
to be consistent with the expectation based on 
Gamow-Teller transitions.
For the isoscalar part of $g_1$, 
it would be interesting to see whether the collapse in 
$g_1^{(p+n)}$ at small $x$ persists at finite nuclear density.
A priori, different contributions to resolving 
the proton spin puzzle (pion cloud, polarized glue) will
come with different $A$ dependence,
e.g. gluons do not directly couple to the meson mean-fields
in the nucleus in the QMC approach,
so any cancellation 
which works for free nucleons might break down at finite density.

\section{The GDH sum-rule in medium}

One also expects medium dependence of the GDH sum-rule and 
the spin-dependent photoabsorption cross-sections with
polarized real photon scattering, $Q^2=0$.
The GDH sum-rule for polarized photon-proton scattering reads 
\cite{Gerasimov:1965et,Drell:1966jv}
\begin{equation}
\int^{\infty}_{M^2} 
\frac{ds_{\gamma p}}{s_{\gamma p} - M^2}
(\sigma_P - \sigma_A)
= 2 \pi^2 \alpha_{\rm QED} \kappa^2 / M^2
\end{equation}
where $\sigma_P$ and $\sigma_A$ 
are the spin dependent photoabsorption cross-sections,
$s_{\gamma p}$ 
is the photon-proton centre of mass energy squared
with $\kappa$ the target's anomalous magnetic moment and
$M$ the target mass.
For free protons with $\kappa=1.79$ 
the sum-rule predicts a value of $205 \ \mu$b
whereas the current value extracted from experiments is
$211 \pm 13 \ \mu {\rm b}$ \cite{Bass:2018uon}.
The dominant contribution to the GDH sum-rule comes from
the $\Delta$ resonance excitation \cite{Helbing:2006zp}
with other resonance contributions averaging to about zero.
There is a 
$\sim 10\%$ 
high-energy Regge contribution in the isovector channel
with negligible isoscalar contribution from centre of mass
energy greater than about 2.5 GeV \cite{Bass:2018uon}.

Both sides of the GDH sum-rule are expected to be enhanced
in medium.
The nucleon and $\Delta$ effective 
masses and the nucleon magnetic 
moments are expected to change in nuclei.
Consider a polarized proton in symmetric nuclear matter.
In the QMC model
the difference in nucleon and $\Delta$ masses, 
$M_N - M_{\Delta}$, is taken as density independent,
with the nucleon mass decreasing
by a factor of $(1-0.2 \rho/\rho_0)$ 
where $\rho$ is the nuclear density \cite{Saito:2005rv}.
Within the same model 
the nucleon magnetic moments increase 
by factor of $(1+0.1 \rho/\rho_0)$ with 
$\mu^*_N / \mu_N 
 \sim
g_A^{(3)}/g_A^{* (3)}$
\cite{Saito:1994kg}.
That is, 
the proton and $\Delta$ resonance masses decrease in medium
whereas the proton magnetic moment increases with increasing nuclear density. 
For the GDH integral, Eq.~(7),
the $\Delta$ resonance contribution to the integral 
will be enhanced at smaller effective $\Delta$ mass, 
weighted by 1/(the incident photon energy in the LAB frame).
Taking the QMC values for the proton effective mass and magnetic moment in medium, 
one finds an enhancement in the GDH integral
by factor of 2.1 at $\rho_0$.
As an independent estimate, 
the effective mass of anti-protons is observed 
in heavy ion collisions 
to be reduced by about 100-150 MeV at density 
$2 \rho_0$~\cite{Schroter:1994ck}.
Making the usual linear density approximation for this 
(anti-)proton effective mass reduction, 
combined with a 20\% reduction in $g_A^{*(3)}$
while still assuming $\mu^*_N / \mu_N 
 \sim
g_A^{(3)}/g_A^{* (3)}$, gives an enhancement in the GDH
integral of factor of 2.2 at $\rho_0$, 
similar to the QMC estimate for this quantity.

\section{Conclusions}

Quenching of the nucleon's axial charge in medium revealed 
in Gamow-Teller transitions provides a sum-rule constraint
on the nucleon's spin structure in medium.
Pions in nuclei
and the Ericson-Ericson-Lorentz-Lorenz effect
modify $g_A^{(3)}$ and 
the partonic spin structure of nucleons in nuclei relative 
to free nucleons.  
Models of the medium dependence of parton structure 
based only on short range nucleon correlations are disfavoured
with SRCs acting to suppress the quenching of $g_A^{(3)}$. 
While a 20\% suppression of the Bjorken sum-rule of 
polarized deep inelastic scsttering is expected 
when scattering from
nucleons in nuclei at nuclear matter density, 
a much larger effect
- factor of two enhancement - is expected in the
Gerasimov-Drell-Hearn sum-rule for polarized photoproduction.
Future experimental study of the GDH sum-rule in medium
would be very interesting
and complement 
deep inelastic measurements of QCD spin effects in nuclei.
Studies of polarized photoproduction with nuclear targets 
might be possible in future experiments at Jefferson Laboratory
\cite{dalton}.

\section*{Acknowledgments}

\noindent
I thank C. Aidala for helpful discussion.


\end{document}